\documentclass[a4paper,11pt]{article}

\usepackage{jcappub}
\usepackage{graphicx}
\usepackage{amsmath,amssymb,amsfonts}
\usepackage{bm}
\usepackage{array,tabularx,multirow,booktabs}
\usepackage{adjustbox}
\usepackage{tikz}
\usepackage{orcidlink}
\usepackage[normalem]{ulem}
\usepackage{xcolor}
\usepackage{microtype}
\usepackage{hyperref}
\hypersetup{colorlinks=true,linkcolor=blue,urlcolor=blue,citecolor=blue}
\allowdisplaybreaks
\UseRawInputEncoding

\arxivnumber{}

\title{Can finite-density QCD matter support a regular black-hole core?}

\author[a,b]{G.~Lambiase \orcidlink{0000-0001-7574-2330}}
\emailAdd{lambiase@sa.infn.it}
\affiliation[a]{Dipartimento di Fisica ``E.R. Caianiello'', Universit\`a degli Studi di Salerno, Via Giovanni Paolo II, 132, 84084 Fisciano (SA), Italy}
\affiliation[b]{Istituto Nazionale di Fisica Nucleare, Gruppo Collegato di Salerno, Sezione di Napoli, Via Giovanni Paolo II, 132, 84084 Fisciano (SA), Italy}

\author[c]{A.~\"Ovg\"un \orcidlink{0000-0002-9889-342X}}
\affiliation[c]{Physics Department, Eastern Mediterranean University, Famagusta, 99628 North Cyprus, via Mersin 10, T\"urkiye}

\author[d,e,f,g,h]{V.~Vertogradov \orcidlink{0000-0002-5096-7696}}
\affiliation[d]{Wilczek Quantum Center, Shanghai Institute for Advanced Studies, Shanghai 201315, China}
\affiliation[e]{University of Science and Technology of China, Hefei 230026, China}
\affiliation[f]{Physics Department, Herzen State Pedagogical University of Russia, 48 Moika Emb., Saint Petersburg 191186, Russia}
\affiliation[g]{Center for Theoretical Physics, Khazar University, 41 Mehseti Street, Baku AZ-1096, Azerbaijan}
\affiliation[h]{SPB Branch of SAO RAS, 65 Pulkovskoe Rd, Saint Petersburg 196140, Russia}

\abstract{
We investigate, within a generalized Vaidya anisotropic reconstruction, whether two finite-density QCD-inspired effective equations of state can generate the near-center stress-energy scaling required for a regular black-hole interior during spherical collapse. The Einstein equations for this geometry fix the radial principal pressure to $P_{\parallel}=-\rho$. We therefore adopt the phenomenological anisotropic closure $P_{\perp}=p_{\rm QCD}$, where $P_{\perp}=-M''/r$ is the transverse principal pressure. This identification should not be interpreted as an isotropic equilibrium description of QCD matter. The resulting identity
$r\rho'+2(\rho+P_{\perp})=0$
determines the radial temperature profile and allows the mass function to be reconstructed without assuming a regular profile a priori. For the chiral model, the conservation equation admits an exact Lambert-$W$ solution, whose high-temperature branch gives
$T^2\propto r^{-296/159}$
and a non-integrable central density. For the finite-density mean-field QGP model, the exact implicit solution gives
$T^2\propto r^{-2\delta/\gamma}$
and
$M-M_0\propto r^{3-4\delta/\gamma}$,
which is incompatible with the regularity condition
$M=\mathcal O(r^3)$
throughout the physical parameter range of the adopted approximation. More generally, if
$P_{\perp}/\rho\to w_0$,
then the generalized Vaidya identity gives
$M\propto r^{1-2w_0}$. Under the transverse null energy condition, a regular center requires the unique limiting behavior
$w_0=-1$. Neither effective QCD closure considered here approaches this vacuum-like limit. Our result is therefore a restricted no-go statement for the generalized Vaidya anisotropic reconstruction: finite quark chemical potential modifies the transition thermodynamics but does not by itself generate a self-regularizing core. We illustrate the additional short-distance physics required for regularization through a smooth compact-support matching to an inner vacuum-like component.
}


\begin{document}
\maketitle
\flushbottom
\tableofcontents
\section{Introduction}

The modern era of black-hole physics is driven simultaneously by gravitational-wave detections and horizon-scale imaging. The first direct observation of gravitational waves from a binary black-hole merger by LIGO/Virgo~\cite{LIGOScientific:2016aoc} and the subsequent images of the supermassive black holes M87* and Sgr~A* by the Event Horizon Telescope (EHT)~\cite{Falcke:1999pj,EventHorizonTelescope:2019dse,EventHorizonTelescope:2022wkp} have transformed black holes from purely theoretical constructs into directly testable astrophysical objects. These observations enable increasingly sharp constraints on deviations from the Kerr paradigm through shadows, strong lensing, ringdown spectra, and stellar dynamics~\cite{Bambi:2019tjh,Vagnozzi:2019apd,Allahyari:2019jqz,Zakharov:2018awx,Zakharov:2018cbj,Zakharov:2021cgx,Vagnozzi:2022moj,Kuang:2022ojj,Kuang:2022xjp,Kuang:2024ugn}. Recent studies have further emphasized that the regularity of black-hole and effective geometries is tightly connected with the structure of the matter sector and the status of the energy conditions, both in broad families of regular spacetimes and in static, spherically symmetric effective geometries~\cite{Wang:2026sqr,Wang:2026jvo}. 
Related developments in modified and effective gravitational frameworks, including Lorentzian--Euclidean black holes, quantum-corrected Schwarzschild geometries, wormhole solutions, and extended geometric formulations of gravity, have shown that causal structure, shadows, accretion, atemporality, and singularity avoidance provide complementary probes of nonclassical compact-object interiors~\cite{Battista:2026gvo,Battista:2026nsx,Capozziello:2025wwl,DeBianchi:2025bgn,Wang:2025fmz,Battista:2024gud,Capozziello:2024ucm,DeFalco:2024ojf,Battista:2023iyu}. In parallel, a large body of work has shown how black-hole shadows, photon spheres, and gravitational lensing encode information about the underlying geometry and matter sector~\cite{Pantig:2025deu,Ovgun:2026ind,Pantig:2026xjj,Perlick:2015vta,Claudel:2000yi,Virbhadra:1999nm,Virbhadra:2007kw,Virbhadra:2022iiy,Virbhadra:2024xpk,Adler:2022qtb,Zhang:2019glo,Luo:2023ndw,Vertogradov:2024dpa,Vertogradov:2024qpf,Shaikh:2019hbm,Dey:2019fpv,Meng:2022kjs,Ling:2021vgk}.

On the theoretical side, the standard general-relativistic picture of gravitational collapse, pioneered by Oppenheimer and Snyder and by Datt~\cite{Oppenheimer:1939ue,datt} and generalized by Penrose, Hawking, and others~\cite{Penrose:1964wq,Hawking:1970zqf,Penrose:1969pc}, predicts the formation of spacetime singularities under broad conditions. This naturally raises the question of whether realistic collapse can instead lead to non-singular compact objects whose exterior resembles a black hole while curvature remains finite in the interior. Regular black holes (RBHs) provide an appealing phenomenological framework for addressing this issue~\cite{Bardeen1968qtr..conf...87B,Dymnikova:1992ux,Ansoldi:2008jw,Lan:2023cvz,Bambi:2023try}. Many RBH metrics approach Schwarzschild or Reissner--Nordstr\"om at large radius but develop a de~Sitter-like or vacuum-like core~\cite{Sakharov1966JETP...22..241S,Gliner1966JETP...22..378G}, often supported by nonlinear electrodynamics or other effective matter sources~\cite{Ayon-Beato:1998hmi,Ayon-Beato:1999qin,Ayon-Beato:1999kuh,Bronnikov:2000vy,Bronnikov:2006fu,Bronnikov:2017sgg,Fan:2016hvf,Fan:2016rih,Barros:2020ghz,Ayon-Beato:2000mjt,Dymnikova:2015yma,Singh:2022xgi,Sudhanshu:2024wqb,Singh:2022ycn,Nicolini:2005vd,Culetu:2022otf,Hayward:2005gi,Markovic:1999di,Hossenfelder:2009fc}.

A key conceptual question is whether such regular geometries can arise dynamically from realistic high-density equations of state rather than being imposed by hand. This issue has been studied from several perspectives, including semiclassical shells and dust collapse, quantum-inspired matter sources, and explicit collapse models involving baryonic phase transitions~\cite{Casadio:1995qy,Alberghi:1998xe,Hossenfelder:2009fc,Baccetti:2018qrp,Buoninfante:2024oyi,Malafarina:2016yuf,Tavakoli:2013tpa,Harko:2013sea,Malafarina:2022oka,Vertogradov:2024seh,Vertogradov:2025yto,Vertogradov:2025snh,Vertogradov:2025aok,Vertogradov:2025jxp}. These studies suggest that regular interiors, if they occur, require either nontrivial effective matter, quantum corrections, or an inner phase with approximately vacuum-like properties.

An essential ingredient in this discussion is the microphysics of matter at extreme densities and temperatures. In QCD, the transition from hadronic matter to a quark--gluon plasma (QGP) has long been connected to early-Universe physics, compact stars, and gravitational-wave signatures~\cite{Capozziello:2018qjs,Sanches:2014gfa,Starobinsky:1979ty}. At small quark chemical potential the finite-temperature QCD transition is generally understood as a crossover, while a first-order transition may arise only in a sufficiently high-density regime. This makes finite chemical potential a natural parameter to explore in collapse models involving extremely dense matter. At the same time, quark matter and strange-star equations of state provide a complementary laboratory for strong anisotropy, large pressure gradients, and modified compact-object structure~\cite{Witten:1984rs,Farhi:1984qu,Cheng:1995am,Dai:1995uj,Cheng:1998na,Ozel:2016oaf,Deb:2016lvi,Rahaman:2014waa,Panotopoulos:2017eig,Panotopoulos:2017pgv,Lopes:2019psm,Panotopoulos:2019zxv,Panotopoulos:2019wsy,Panotopoulos:2020zqa,Panotopoulos:2017jdc,Panotopoulos:2020uvq,Panotopoulos:2024jtn,Harko:2000ni,Yavuz:2005qb,Olmo:2019flu}.

In this work we ask a deliberately restricted question: can the two adopted finite-density QCD-inspired effective equations of state generate the near-center scaling required for regularity when embedded in the generalized Vaidya geometry? In this geometry the Einstein equations impose
$P_{\parallel}=-\rho$,
so an isotropic QCD perfect fluid cannot be inserted without modification. We instead use the phenomenological anisotropic closure
$P_{\perp}=p_{\rm QCD}$.
The conclusions derived below therefore apply to this generalized Vaidya reconstruction and should not be interpreted as a general no-go theorem for all possible dynamical configurations of QCD matter. The analysis is an effective-fluid reconstruction and not a first-principles simulation of nonequilibrium QCD collapse. Following Ref.~\cite{Capozziello:2018qjs}, we first review an effective three-flavor chiral quark model at finite $(T,\mu)$ that yields an improved QGP equation of state and an explicit expression for the critical temperature $T_c(\mu)$. We then complement this description with a cold-QGP relativistic mean-field model~\cite{Sanches:2014gfa}, in which the gluon field is split into soft and hard momentum modes and the hard modes acquire a dynamical mass. On the geometric side, we reconstruct the mass function $M(v,r)$ directly from local energy-momentum conservation in an advanced Eddington--Finkelstein gauge, rather than postulating a regular mass profile from the outset.

The analysis leads to a clear conditional conclusion. Within the generalized Vaidya ansatz and the phenomenological identification
$P_{\perp}=p_{\rm QCD}$,
neither of the two adopted effective QCD closures produces the near-center scaling required for a regular core. In the chiral model, the exact conservation equation is solved in terms of the Lambert $W$ function, but the physical high-temperature branch is singular. In the finite-density mean-field QGP model, the conservation equation drives the temperature to diverge toward the center, and the mass function fails to exhibit the de~Sitter-like behavior
$M\sim r^3$.
The precise numerical exponents depend on the adopted effective thermodynamic pairs, whereas the qualitative obstruction follows more generally whenever the limiting transverse ratio satisfies
$P_{\perp}/\rho>-1$.
A regular completion therefore requires an additional short-distance sector whose stress tensor becomes approximately vacuum-like near the center.

The paper is organized as follows. In Sec.~\ref{sec:chiralEoS} we briefly review the effective chiral QCD model with finite chemical potential and extract the corresponding QGP EoS in a form suitable for our purposes. In Sec.~\ref{sec:BH-chiral} we embed this EoS into the Eddington--Finkelstein (EF) geometry, derive the exact temperature profile in terms of the Lambert $W$ function, and analyze the near-center behavior of the resulting mass function. Section~\ref{sec:coldQGP} is devoted to the cold-QGP mean-field model, where we derive the relevant EoS and discuss the $\mu$-dependent critical temperature. In Sec.~\ref{sec:BH-coldQGP} we solve the conservation law for the cold-QGP EoS, obtain an exact implicit $r$--$T$ relation, and analyze the regularity of the resulting mass function. Section~\ref{sec:transition} discusses a finite-time phenomenological conversion of baryonic matter into quark matter and radiation, and demonstrates that a positive-energy baryon--radiation mixture cannot by itself generate the vacuum-like pressure required for a de~Sitter core. In Sec. \ref{sec:corematching}, we study explicit matching to an inner vacuum-like core. We summarize our conclusions and comment on observational prospects---including implications for black hole shadows, lensing, and quasinormal spectra in light of current and forthcoming EHT and gravitational-wave observations~\cite{Falcke:1999pj,EventHorizonTelescope:2019dse,EventHorizonTelescope:2022wkp,LIGOScientific:2016aoc,Vagnozzi:2022moj,Kuang:2022ojj,Kuang:2022xjp,Lambiase:2024uzy,Lambiase:2024lvo,Pantig:2024asu}---in Sec.~\ref{sec:conclusion}.

\section{Brief review of a QCD effective model with finite chemical potential}
\label{sec:chiralEoS}

To make the discussion self-contained, we first summarize the three-flavor chiral quark model with finite chemical potential introduced in Ref.~\cite{Capozziello:2018qjs}. The main idea is to encode the QCD dynamics near a first-order phase transition in an effective scalar-pseudoscalar chiral field coupled to quarks, and to derive an improved EoS for the QGP phase, including the dependence on the quark chemical potential $\mu$.

At the thermodynamic level, the matter content in the quark-gluon plasma (QGP) and hadron phases is described by the equations of state
\begin{equation}
p_{QGP}=\frac{g_{QGP}\pi^2}{90}T^4 - V(T)\,, 
\qquad
\rho_{QGP}=\frac{g_{QGP}\pi^2}{30}T^4 + V(T)\,,
\end{equation}
and
\begin{equation}
 \label{eq:hadron-radiation}
 p_{H}=\frac{\rho_{H}}{3}=\frac{g_{H}\pi^2}{90}T^4\,,
\end{equation}
where $p$ is the pressure, $\rho$ the energy density, and $T$ the temperature. The functions $g_{QGP}$ and $g_H$ denote the effective number of relativistic degrees of freedom in the QGP and hadron phases, respectively, while $V(T)$ encodes self-interaction effects (e.g.\ the bag contribution) in the QGP phase. The hadronic phase is modeled as a relativistic gas with the conformal relation $p_H = \rho_H/3$.

This assumption is appropriate for an ultrarelativistic hadronic component. At sufficiently high baryon density, a Zel'dovich-type stiff phase with
$p_H=\rho_H$
may instead become relevant. Such a phase cannot be introduced consistently by retaining
$\rho_H\propto T^4$
and merely replacing
$p_H=\rho_H/3$
by
$p_H=\rho_H$.

At finite baryon chemical potential $\mu_B$, thermodynamic consistency requires
\begin{equation}
\rho_H
=
-p_H
+
T\left(\frac{\partial p_H}{\partial T}\right)_{\mu_B}
+
\mu_B\left(\frac{\partial p_H}{\partial\mu_B}\right)_T.
\label{eq:hadron-thermo-general}
\end{equation}
Consequently, a realistic high-density stiff phase must be specified by a two-variable free energy or pressure
$p_H(T,\mu_B)$.

Only in the special limit of vanishing or neglected hadronic chemical potential does
\begin{equation}
\rho_H=T\frac{dp_H}{dT}-p_H
\end{equation}
hold. In that restricted limit, a constant barotropic index
$p_H=w_H\rho_H$
gives
\begin{equation}
p_H(T)=K_H T^{(1+w_H)/w_H},
\qquad
\rho_H(T)=\frac{p_H(T)}{w_H}.
\label{eq:hadron-consistent-w}
\end{equation}
Thus the radiation case
$w_H=1/3$
scales as $T^4$, whereas the formal zero-$\mu_B$ stiff limit
$w_H=1$
scales as $T^2$. A quantitative stiff-matter transition curve requires an independent high-density hadronic free-energy model. This uncertainty can change the position and structure of the transition layer, but it does not affect the conditional near-center result once the system is assumed to be on one of the QGP branches analyzed below.

These equations follow from the effective Lagrangian density
\begin{equation}
\mathcal{L} = \sum_{k=1}^{n_{ f}} \left[ i\bar{\psi}_k \gamma^\mu  \partial_\mu \psi_k 
- g \bar{\psi}_k (\sigma + i \tau \cdot \pi  \gamma_5) \psi_k \right] 
+  \frac{1}{2} \partial_\mu \sigma \partial^\mu \sigma
+\frac{1}{2}\partial_\mu \pi \partial^\mu \pi - V(\sigma^2 + \pi^2)\,,
\end{equation}
where $\psi_k$ are the quark fields, $\sigma$ is a scalar meson field, and $\pi$ denotes the pion triplet. The Yukawa coupling $g$ controls the quark mass generated by chiral symmetry breaking. The self-interaction potential $V$ is responsible for the shape of the chiral effective potential and the existence of (meta)stable minima.

It is convenient to rewrite the Lagrangian in terms of the chiral radius
\(
\varphi = (\sigma^2 + \pi^2)^{1/2}
\)
and a unitary matrix $U$ describing the direction of the chiral field in isospin space. In this parametrization one finds
\begin{equation}
{\cal L} = \sum_{k=1}^{n_{\rm f}} \left[ i\bar{\psi}_k \gamma^\mu  \partial_\mu \psi_k 
- g \varphi (\bar{\psi}_k^{\rm L} U \psi_k^{\rm R} +  \bar{\psi}_k^{\rm R} U^\dagger \psi_k^{\rm L}) \right]
+\frac{1}{2} \partial_\mu  \varphi \partial^\mu \varphi + \frac{1}{4} \varphi^2\, \mathrm{Tr} (\partial_\mu U \partial^\mu U^\dagger) - V(\varphi)\,.
\end{equation}
Here $\psi^{\rm L,R}$ denote left- and right-handed quark fields, and $V(\varphi)$ depends only on the magnitude of the chiral field and acts as an order-parameter potential. The dynamics of $\varphi$ capture the chiral symmetry breaking and restoration.

The self-interaction potential is chosen as
\begin{equation}
V(\varphi)=\frac{1}{2} f_\pi^2 \left(\lambda^2  - \frac{12B}{f_\pi^4}\right)  \varphi^2 \left(1  - \frac{\varphi}{f_\pi}\right)^2
+ B \left[1 + 3\left(\frac{\varphi}{f_\pi}\right)^4 - 4\left(\frac{\varphi}{f_\pi}\right)^3 \right]\,,
\end{equation}
where $f_\pi$ is the pion decay constant and $B$ represents the vacuum-energy difference between the perturbative and nonperturbative phases. For the parameter range employed in the effective model, $\varphi=f_\pi$ represents the broken-symmetry vacuum, whereas $\varphi=0$ can represent a metastable chirally restored configuration. The existence and ordering of these extrema depend on the model parameters and should not be inferred solely from the normalization of the potential.

The potential is therefore not introduced merely as an algebraic interpolation. It is a phenomenological chiral-bag/linear-sigma effective potential intended to encode the order parameter and the vacuum-energy difference between the competing QCD phases. Its normalization gives
\begin{equation}
V(0)=B,
\qquad
V(f_\pi)=0.
\end{equation}
The parameters $B$, $f_\pi$, and $\lambda$ control the false-vacuum energy, the location of the broken-symmetry configuration, and the curvature and barrier structure. The restored configuration at $\varphi=0$ is locally stable when
\begin{equation}
V''(0)
=
f_\pi^2
\left(
\lambda^2-\frac{12B}{f_\pi^4}
\right)>0.
\label{eq:potential-local-minimum}
\end{equation}
For $B>0$ and parameter values for which no lower additional extremum is present, the configuration at $\varphi=f_\pi$ is energetically preferred over $\varphi=0$. We stress that this potential is an effective phenomenological model and not a first-principles derivation from nonperturbative QCD.

To incorporate finite quark chemical potential $\mu$ and temperature $T$, one adds the fermionic contribution to the thermodynamic potential,
\begin{equation}
\omega_f(T, \mu) = -T\left[\int\!\frac{d^3k}{(2\pi)^3} \ln\left(1 + e^{-(E(k)-\mu)/T}\right)
+  \int\!\frac{d^3k}{(2\pi)^3} \ln\left(1 + e^{-(E(k)+\mu)/T}\right)\right]\,,
\end{equation}
where $E(k)=\sqrt{k^2+m_s^2}$ for the strange quark (here $m_s$ is the mass of strange quark); the up and down quarks are treated as effectively massless. The first and second integrals represent the contributions of quarks and antiquarks, respectively. Expanding $\omega_f$ in powers of $T$ and $\mu$ for a non-zero strange quark mass $m_s$ yields corrections to the effective potential:
\begin{equation}
V_{T, \mu}(\varphi) = B - \alpha_T T^4 -\frac{3\mu^2}{2} T^2 - \alpha_\mu \mu^4 + \gamma_T T^2 + \gamma_\mu \mu^2\,,
\end{equation}
with
\begin{equation}
\alpha_T = \frac{7\pi^2}{20}, \qquad \alpha_\mu = \frac{3}{4\pi^2}, \qquad 
\gamma_T = \frac{m_s^2}{4}, \qquad \gamma_\mu = \frac{3m_s^2}{4\pi^2}\,.
\end{equation}
These coefficients encode thermal and density corrections to the vacuum energy and effectively renormalize the bag parameter and the curvature of the potential.

Combining the above ingredients, one arrives at the improved EoS for the QGP phase:
\begin{align}
\label{p_QGP}
p_{QGP} &= \left(\frac{37\pi^2}{90}+ \alpha_T\right)T^4 +\left(\frac{3\mu^2}{2}-\gamma_T \right)T^2 + \alpha_\mu \mu^4 - \gamma_\mu \mu^2 - B\,, \\
\label{rho_QGP}
\rho_{QGP}&= \left(\frac{37\pi^2}{30}- \alpha_T\right)T^4 -\left(\frac{3\mu^2}{2}-\gamma_T \right)T^2 - \alpha_\mu \mu^4 + \gamma_\mu \mu^2 + B\,.
\end{align}
The $T^4$ terms are dominated by the relativistic quark and gluon degrees of freedom, while the $T^2$ and $\mu^4$ terms encode the effects of the strange quark mass and finite baryon density, respectively. In the limit $\alpha_T=\gamma_T=\alpha_\mu=\gamma_\mu=0$ one recovers the standard MIT bag relation $p=\frac{1}{3}(\rho-4B)$.

The critical temperature $T_c$ of the phase transition is obtained by imposing pressure equality between the QGP and hadron phases, $p_{QGP}(T_c,\mu)=p_H(T_c)$. Using the above EoS one finds
\begin{equation}
\label{critical}
T_{c}=\left[\frac{\frac{3\mu^2}{2}-\gamma_T }{\frac{\pi^2}{45}\Delta g+ 2\alpha_T}\, 
\left(-1\pm \sqrt{1-\frac{\left(\frac{2\pi^2}{45}\Delta g+ 4\alpha_T\right)\left(\alpha_\mu \mu^4 - \gamma_\mu \mu^2-B \right)}{\left(\frac{3\mu^2}{2}-\gamma_T \right)^2}}\right)\right]^{1/2}\,,
\end{equation}
where $\Delta g = g_{QGP}-g_H$ is the jump in the effective number of relativistic degrees of freedom across the transition. The physically relevant branch corresponds to the choice of sign that yields a real and positive $T_c$. The dependence of $T_c$ on $\mu$ will be important later when we consider collapse scenarios with non-negligible quark chemical potential. 

The pressure-equality equation above applies only to the radiation-like hadronic reference phase. A finite-density stiff phase must instead be specified by a pressure
$p_H(T,\mu_B)$,
and the phase boundary must be obtained from
\begin{equation}
p_{\rm QGP}(T_c,\mu)
=
p_H(T_c,\mu_B),
\label{eq:general-phase-equilibrium}
\end{equation}
together with the relevant chemical-equilibrium relation. For equal quark chemical potentials one has
$\mu_B=3\mu$.
In general, the resulting transition equation is not quadratic in $T_c^2$. We therefore do not implement a purely algebraic replacement of the radiation equation of state by a constant stiff index in the numerical analysis.

A particularly useful combination of the above equations is
\begin{equation}
\label{eq:eos1}
p_{QGP}=-\,\rho_{QGP}+\frac{74\pi^2}{45}T^4\,.
\end{equation}
This relation shows that, within this model, the deviation from a vacuum-like equation of state $p=-\rho$ is controlled entirely by the $T^4$ term, while the bag and chemical-potential contributions cancel in $p+\rho$. This structure will play a key role when we embed the EoS into a dynamical black hole geometry.

We emphasize the phenomenological status of Eqs.~\eqref{p_QGP}--\eqref{rho_QGP}. In the present work the quoted pressure and energy density are treated as the independent effective closure supplied by Ref.~\cite{Capozziello:2018qjs}; they are not rederived from a single grand potential using
\begin{equation}
\rho
=
-p
+
T\left(\frac{\partial p}{\partial T}\right)_{\mu}
+
\mu\left(\frac{\partial p}{\partial\mu}\right)_T.
\end{equation}
The exact numerical exponent
$296/159$
obtained below is therefore a property of this adopted phenomenological pair. A thermodynamically closed equation of state may modify the numerical coefficients and the exact exponent. The more general conclusion remains valid whenever the high-density transverse equation of state approaches a finite ratio
$P_{\perp}/\rho=w_0>-1$,
because the generalized Vaidya identity then gives a mass exponent strictly smaller than three. A first-principles treatment would have to evolve the conserved quark density, chemical potential, temperature, and stress anisotropy dynamically.

Before discussing special limits, it is important to stress that the physical coefficients of the chiral model already fix the asymptotic exponent. Using
\begin{equation} \label{alphaxizeta}
\alpha_T=\frac{7\pi^2}{20},
\qquad
\xi=\frac{37\pi^2}{30}-\alpha_T=\frac{53\pi^2}{60},
\qquad
\zeta=\frac{74\pi^2}{45},
\end{equation}
one obtains
\begin{equation}
\frac{\zeta}{\xi}=\frac{296}{159}\approx 1.8616,
\qquad
\frac{2\zeta}{\xi}=\frac{592}{159}\approx 3.7233.   \label{xizet}
\end{equation}

\section{Chiral QCD equation of state in the Eddington--Finkelstein geometry}
\label{sec:BH-chiral}

We now couple the chiral QCD equation of state discussed in Sec.~\ref{sec:chiralEoS} to a dynamical, spherically symmetric geometry. We use advanced Eddington--Finkelstein coordinates,
\begin{equation}
\label{eq:metric-general}
ds^2=-\left(1-\frac{2M(v,r)}{r}\right)dv^2+2\,dv\,dr+r^2d\Omega^2,
\end{equation}
where $M(v,r)$ is the Misner--Sharp mass function. Throughout the gravitational reconstruction we use units $8\pi G=c=\hbar=k_{\rm B}=1$.

For the generalized Vaidya geometry, the nonvanishing mixed Einstein-tensor components are
\begin{equation}
G^v{}_v=G^r{}_r=-\frac{2M'}{r^2},
\qquad
G^r{}_v=\frac{2\dot M}{r^2},
\qquad
G^\theta{}_\theta=G^\varphi{}_\varphi=-\frac{M''}{r}.
\label{eq:einstein-mixed}
\end{equation}
The compatible type-II effective stress tensor can be written as
\begin{equation}
T_{\mu\nu}=\sigma l_\mu l_\nu+(\rho+P_\perp)
(l_\mu n_\nu+l_\nu n_\mu)+P_\perp g_{\mu\nu},
\label{eq:Tmunu-typeII}
\end{equation}
where $l_\mu l^\mu=n_\mu n^\mu=0$ and $l_\mu n^\mu=-1$. The Einstein equations give
\begin{equation}
\sigma=\frac{2\dot M}{r^2},
\qquad
\rho=\frac{2M'}{r^2},
\qquad
P_\perp=-\frac{M''}{r},
\qquad
P_\parallel=-\rho.
\label{eq:stress-identification}
\end{equation}
Thus $-M''/r$ is the transverse principal pressure, not the radial principal pressure. The radial pressure is fixed geometrically to $P_\parallel=-\rho$. The identity following from Eq.~\eqref{eq:stress-identification} is
\begin{equation}
r\rho'+2(\rho+P_\perp)=0.
\label{eq:cons-law-new}
\end{equation}
In the phenomenological reconstruction below, we adopt
\begin{equation}
P_\perp=p_{\rm QCD}.
\label{eq:qcd-transverse-identification}
\end{equation}
This is an anisotropic effective closure motivated by the tensor structure of the generalized Vaidya geometry. It is not an identification of an isotropic equilibrium QCD fluid with the complete stress tensor, because the geometry independently fixes
\begin{equation}
P_\parallel=-\rho.
\end{equation}
Accordingly, the no-go result established below applies to the generalized Vaidya reconstruction with
$P_\perp=p_{\rm QCD}$,
rather than to every possible gravitational configuration of finite-density QCD matter.

The mass function is reconstructed from
\begin{equation}
\label{eq:mass-general-chiral}
M(v,r)=M_0(v)+\frac12\int \rho(r)r^2\,dr,
\end{equation}
where $M_0(v)$ is an integration function and should not be interpreted as a finite central mass when the density is non-integrable.

We introduce $u(r)=T^2(r)$ and write the chiral QCD effective pair as
\begin{equation}
\label{eq:abcd}
p_{QGP}=a u^2+b u+c,
\qquad
\rho_{QGP}=d u^2-bu-c,
\end{equation}
with
\begin{equation}
\label{eq:abcd-def}
a=\frac{37\pi^2}{90}+\alpha_T,
\quad
d\equiv\xi=\frac{37\pi^2}{30}-\alpha_T,
\quad
b=\frac{3\mu^2}{2}-\gamma_T,
\quad
c=\alpha_\mu\mu^4-\gamma_\mu\mu^2-B.
\end{equation}
Defining
\begin{equation}
\eta\equiv-b=\gamma_T-\frac{3\mu^2}{2},
\qquad
\zeta\equiv a+d=\frac{74\pi^2}{45},
\label{eq:eta-zeta}
\end{equation}
we have
\begin{equation}
\rho_{QGP}=\xi u^2+\eta u-c,
\qquad
P_\perp+\rho_{QGP}=p_{QGP}+\rho_{QGP}=\zeta u^2.
\end{equation}
Substitution into Eq.~\eqref{eq:cons-law-new} gives
\begin{equation}
\frac{du}{dr}=-\frac{2\zeta u^2}{r(2\xi u+\eta)}.
\label{eq:udiff-chiral}
\end{equation}

To make the integration constant and the argument of the Lambert function dimensionally transparent, we impose the boundary data
\begin{equation}
u(r_0)=u_0=T_0^2,
\qquad
y_0\equiv\frac{\eta}{2\xi u_0}.
\end{equation}
Subtracting the integrated equation at $(r_0,u_0)$ gives
\begin{equation}
2\xi\ln\!\left(\frac{u}{u_0}\right)
-\eta\left(\frac1u-\frac1{u_0}\right)
=-2\zeta\ln\!\left(\frac r{r_0}\right).
\label{eq:implicit-u-new}
\end{equation}
The exact solution is therefore
\begin{equation}
\label{eq:uLambert}
u(r)=T^2(r)=
\frac{\eta}{2\xi W\!\left[y_0e^{y_0}(r/r_0)^{\zeta/\xi}\right]}.
\end{equation}

\begin{figure}[tb]
\centering
\includegraphics[width=\textwidth]{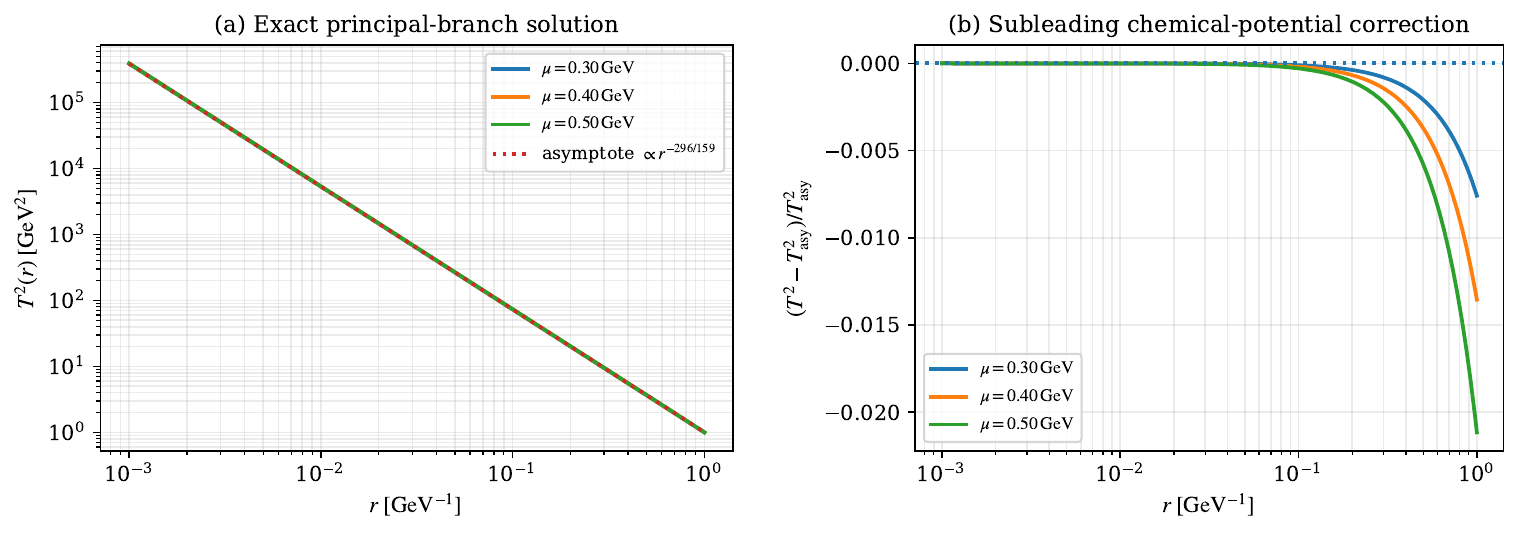}
\caption{Principal-branch Lambert-$W$ temperature profile with the explicit boundary condition $T(r_0)=T_0$, where $r_0=1~{\rm GeV}^{-1}$ and $T_0=1~{\rm GeV}$. Panel~(a): exact solutions for $\mu=0.30,0.40,0.50~{\rm GeV}$ and the universal near-center power $T^2\propto r^{-296/159}$. Panel~(b): fractional difference between the exact solution and the corresponding asymptotic expression. The chemical potential changes only the subleading correction and normalization, not the exponent.}
\label{fig:chiral_T2}
\end{figure}

For the collapse regime one typically has $\eta<0$. For the boundary data used in Fig.~\ref{fig:chiral_T2}, we restrict
\begin{equation}
-1<y_0<0.
\label{eq:Lambert-boundary-domain}
\end{equation}
The Lambert argument
\begin{equation}
z(r)=y_0e^{y_0}\left(\frac r{r_0}\right)^{\zeta/\xi}
\end{equation}
then remains in the real interval $[-e^{-1},0)$ for
$0<r\leq r_0$.
The principal branch satisfies
\begin{equation}
W_0(y_0e^{y_0})=y_0,
\end{equation}
and is therefore the branch that obeys the imposed boundary condition
$u(r_0)=u_0$.
It gives
$T^2\to\infty$
as $r\to0^+$.
The branch $W_{-1}$ corresponds to a distinct low-temperature solution with different boundary data and gives
$T^2\to0$
as $r\to0^+$; it is outside the assumed high-temperature QGP regime.

Since $W_0(z)\sim z$ as $z\to0$, define
\begin{equation}
\mathcal A_\mu\equiv u_0e^{-y_0}.
\end{equation}
Then
\begin{equation}
T^2(r)\sim\mathcal A_\mu\left(\frac r{r_0}\right)^{-\zeta/\xi},
\qquad r\to0^+,
\label{eq:u-small-r}
\end{equation}
and therefore
\begin{equation}
P_\perp+\rho_{QGP}=\zeta T^4
\sim\zeta\mathcal A_\mu^2\left(\frac r{r_0}\right)^{-2\zeta/\xi}.
\label{eq:pplusrho-asymptotic-chiral}
\end{equation}

The transverse stress therefore moves away from the vacuum-like condition $P_\perp=-\rho$ as the center is approached.

The exact density is
\begin{equation}
\rho(r)=\xi u(r)^2+\eta u(r)-c,
\label{eq:rho-exact-chiral}
\end{equation}
and hence
\begin{equation}
M(v,r)=M_0(v)+\frac12\int[\xi u(r)^2+\eta u(r)-c]r^2\,dr.
\label{eq:mass-exact-chiral}
\end{equation}

The leading density behaves as
\begin{equation}
\rho(r)\sim\xi\mathcal A_\mu^2
\left(\frac r{r_0}\right)^{-2\zeta/\xi}.
\label{eq:rho-asymptotic-chiral}
\end{equation}
Thus a function has the asymptotic form
\begin{equation}
M(v,r)\sim M_0(v)+
\frac{\xi\mathcal A_\mu^2r_0^3}{2(3-2\zeta/\xi)}
\left(\frac r{r_0}\right)^{3-2\zeta/\xi}.
\label{eq:mass-asymptotic-chiral-2}
\end{equation}

For the physical coefficients,
\begin{equation}
\frac{\zeta}{\xi}=\frac{296}{159},
\qquad
3-\frac{2\zeta}{\xi}=-\frac{115}{159},
\end{equation}
so
\begin{equation}
M(v,r)\sim M_0(v)+\widehat D\,r^{-115/159},
\qquad \widehat D<0
\label{eq:mass-singular-physical}
\end{equation}
for positive density normalization. More invariantly, $\rho r^2$ is non-integrable at the origin, so no finite Misner--Sharp mass can be defined by integrating from $r=0$. This is incompatible with the regular-center requirement
\begin{equation}
M(v,r)=\mathcal O(r^3).
\end{equation}

\begin{figure}[tb]
\centering
\includegraphics[width=0.6\textwidth]{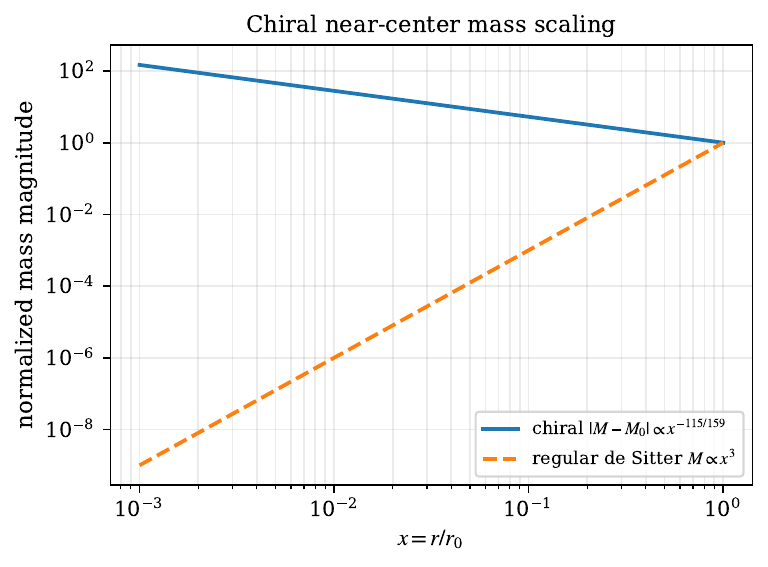}
\caption{Normalized near-center mass scaling in the chiral model. Both curves are normalized at $x=r/r_0=1$. The singular branch obeys $|M-M_0|\propto x^{-115/159}$, whereas a regular de~Sitter core obeys $M\propto x^3$. The physical coefficient of the chiral function  is negative for positive density; the absolute value is plotted only to compare the exponents.}
\label{fig:chiral_mass}
\end{figure}

The chiral QCD effective pair therefore fails to generate a self-regularizing center. Finite chemical potential affects subleading terms and the transition curve, but not the singular leading exponent.

\section{Cold--QGP model and the quark chemical potential}
\label{sec:coldQGP}

We now consider a complementary finite-density mean-field description of quark matter based on the model referred to as the cold-QGP model in Ref.~\cite{Sanches:2014gfa}. The framework separates soft and hard gluon modes and introduces a dynamical gluon mass $m_g$. The analytic equations used below are obtained in the high-temperature limit
$T\gg\mu_f,m_f$.
The terminology should therefore not be interpreted as implying
$T\ll\mu$.
The high-temperature reduction provides a simple effective equation of state that can be analyzed analytically in the generalized Vaidya reconstruction.

The general expressions for the pressure $p$ and energy density $\rho$ in this model are
\begin{align}
p_{\text{cold-QGP}} &= \frac{3\pi\alpha_s}{4{m_{g}}^{2}}{n_q}^{2}-B
+ \sum_{f}\,\frac{\gamma_{f}}{6\pi^{2}}\int_{0}^{\infty} dk \,\frac{k^{4}}{\mathcal{E}_{f}}\, \Big(d_{f}+\bar{d}_{f} \Big)
+ \frac{\gamma_g}{6\pi^{2}}\int_{0}^{\infty} d{k}\,\frac{k^{3}}{e^{k/T}-1}\,, \\
\rho_{\text{cold-QGP}} &= \frac{3\pi\alpha_s}{4{m_{g}}^{2}}{n_q}^{2} + B
+ \sum_{f}\,\frac{\gamma_{f}}{2\pi^{2}} \int_{0}^{\infty} dk \, k^{2}\, \mathcal{E}_{f}\,\Big(d_{f}+ \bar{d}_{f}\Big)
+ \frac{\gamma_g}{2\pi^{2}}\int_{0}^{\infty} d{k}\,\frac{k^{3}}{e^{k/T}-1}.\,
\end{align}
Here $\mathcal{E}_{f}=\sqrt{m_{f}^{2}+k^{2}}$ is the quark single-particle energy, $d_f$ and $\bar{d}_f$ are the Fermi-Dirac distribution functions for quarks and antiquarks with chemical potential $\nu_f$, and $n_q$ is the quark number density. The parameters $\alpha_s$ and $m_g$ represent the strong coupling and the dynamical gluon mass, respectively, while $B$ is the bag constant. The factors $\gamma_f$ and $\gamma_g$ account for spin and color degeneracies of quarks and gluons.

For a system with two light quarks ($u$ and $d$) in the high-temperature regime ($T\gg \nu_{f}$, $T\gg m_{f}$), the integrals can be evaluated analytically. In this limit the EoS simplifies to
\begin{align}
p_{\text{cold-QGP}} &= \left(\frac{37\pi^{2}}{90}+\frac{3\pi\alpha_s\mu^{2}}{4{m_{g}}^{2}}\right)T^{4} + \frac{\mu^{2}}{2} T^{2} - B\,, 
\label{eq:p_simple} \\
\rho_{\text{cold-QGP}}&= \left(\frac{37\pi^{2}}{30}+\frac{3\pi\alpha_s\mu^{2}}{4{m_{g}}^{2}}\right)T^{4} + \frac{3\mu^{2}}{2} T^{2} + B\,, 
\label{eq:rho_simple}
\end{align}
where the chemical potential is taken equal for the two flavors, $\mu \equiv \nu_u = \nu_d$. The first terms in Eqs.~\eqref{eq:p_simple}-\eqref{eq:rho_simple} describe the contribution of relativistic massless quarks and gluons, corrected by an interaction term proportional to $\alpha_s\mu^2/m_g^2$, while the $T^2$ terms originate from finite chemical potential and the bag constant $B$ provides the vacuum energy offset. In the limit $\alpha_s\rightarrow 0$ these expressions reduce to the MIT bag EoS for a two-flavor QGP.

The analytic reduction in Eqs.~\eqref{eq:p_simple}--\eqref{eq:rho_simple} assumes $T\gg\mu_f$ and $T\gg m_f$. Results obtained at $\mu\gtrsim T$ must therefore be regarded as an extrapolation of the high-temperature approximation. As in the chiral case, we use the pressure and density quoted in Ref.~\cite{Sanches:2014gfa} as a phenomenological effective pair. A fully thermodynamically closed treatment would evolve the quark number density and determine $\mu(v,r)$ dynamically.

The critical temperature of the phase transition is obtained by imposing pressure equality between the QGP and hadron phases,
\begin{equation}
p_{\rm cold\text{-}QGP}(T_c,\mu)=p_H(T_c)=\frac{g_H\pi^2}{90}\,T_c^4.
\end{equation}
Using Eq.~\eqref{eq:p_simple}, one finds
\begin{equation}
A_{\rm cold}\,T_c^4+\frac{\mu^2}{2}\,T_c^2-B=0,
\end{equation}
where
\begin{equation}
A_{\rm cold}\equiv \frac{\pi^2}{90}(37-g_H)+\frac{3\pi\alpha_s\mu^2}{4m_g^2}.
\end{equation}
Solving for $T_c^2$, the physically relevant branch is
\begin{equation}
\label{eq:critical-cold}
T_c^2=
\frac{-\mu^2/2+\sqrt{\mu^4/4+4A_{\rm cold}B}}{2A_{\rm cold}}.
\end{equation} 

Equation~\eqref{eq:critical-cold} applies only to the radiation-like hadronic reference phase. For a finite-density stiff phase, one must instead solve
\begin{equation}
p_{\rm cold\text{-}QGP}(T_c,\mu)
=
p_H(T_c,\mu_B)
\end{equation}
using an independently specified high-density hadronic free energy and the appropriate chemical-equilibrium relation. The resulting equation is generally not quadratic in $T_c^2$. We therefore do not use an algebraic substitution
$p_H=w_H\rho_H$
in the numerical transition curve.

Combining Eqs.~\eqref{eq:p_simple} and~\eqref{eq:rho_simple} we obtain
\begin{equation}
\label{eq:coldQGP-rel}
\rho_{\rm cold\text{-}QGP} =3p_{\rm cold\text{-}QGP} +4B -\frac{3\pi\alpha_s\mu^2}{2m_g^2}T^4
\end{equation}
and hence
\begin{equation} \label{eq:eos222}
p_{\rm cold\text{-}QGP} =\frac{1}{3}\rho_{\rm cold\text{-}QGP} -\frac{4B}{3}
+\frac{\pi\alpha_s \mu^2}{2m_g^2}T^4.
\end{equation}
The cold-QGP fluid therefore behaves as a radiation-like fluid $p\simeq \rho/3$ with a negative vacuum contribution $-4B/3$, plus an interaction correction proportional to $\alpha_s\mu^2 T^4/m_g^2$. When $\alpha_s\to0$ we recover exactly the MIT-bag form $p=\tfrac13(\rho-4B)$.

\begin{figure}[tb]
\centering
\includegraphics[width=0.6\textwidth]{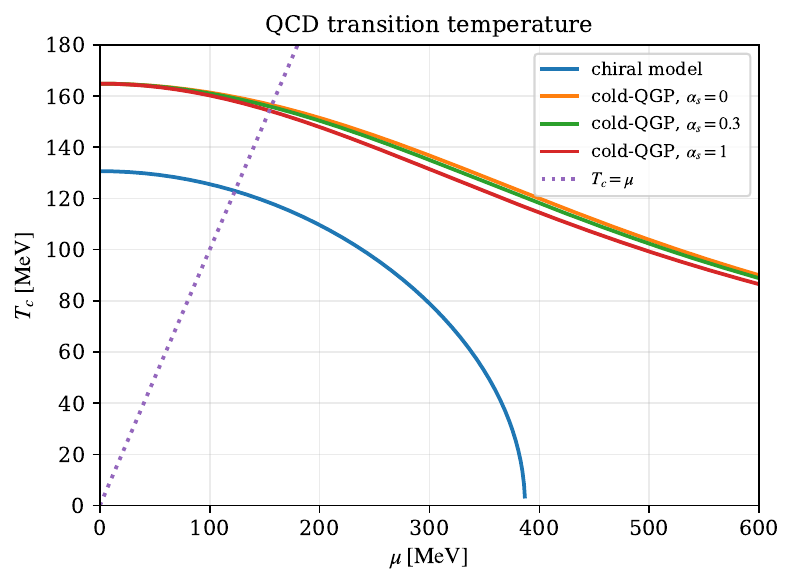}
\caption{Critical temperature $T_c(\mu)$ for the radiation-like hadronic reference phase with $g_H=17.25$. The chiral curve uses $B^{1/4}=200$~MeV and $m_s=95$~MeV; the cold-QGP curves use $m_g=600$~MeV and $\alpha_s=0,0.3,1.0$. Direct solution of the pressure-equality equations gives $T_c(0)=130.67$~MeV for the chiral model and $164.86$~MeV for the cold-QGP model. The chiral branch terminates at $\mu\simeq386.89$~MeV where $T_c\to0$. The dotted diagonal $T_c=\mu$ marks the boundary beyond which even the weaker inequality $T_c>\mu$ fails; the cold-QGP curves below this line are extrapolations outside the stated high-temperature regime $T\gg\mu$.}
\label{fig:Tc}
\end{figure}

Figure~\ref{fig:Tc} illustrates the model dependence of the transition curve. Both models predict a decrease of $T_c$ with increasing $\mu$, but they do not coincide quantitatively even at $\mu=0$. The line $T_c=\mu$ makes the approximation domain explicit: once a cold-QGP curve falls below this line, the analytic reduction used in Eqs.~\eqref{eq:p_simple}--\eqref{eq:rho_simple} is no longer quantitatively controlled.

\section{Cold-QGP equation of state in the Eddington--Finkelstein geometry}
\label{sec:BH-coldQGP}

We now identify the cold-QGP pressure with the transverse pressure $P_\perp$ in Eq.~\eqref{eq:qcd-transverse-identification}. The conservation identity becomes
\begin{equation}
r\rho'_{\rm cold\text{-}QGP}+2\left(\rho_{\rm cold\text{-}QGP}+p_{\rm cold\text{-}QGP}\right)=0.
\label{eq:nepr2}
\end{equation}
Define
\begin{align}
\alpha&\equiv\frac{3\mu^2}{\gamma},
&\beta&\equiv\frac{4\mu^2}{\delta},\\
\gamma&\equiv\frac{74\pi^2}{15}+\frac{3\pi\alpha_s\mu^2}{m_g^2},
&\delta&\equiv\frac{148\pi^2}{45}+\frac{3\pi\alpha_s\mu^2}{m_g^2}.
\label{eq:defcon2}
\end{align}
Then
\begin{equation}
\gamma(T^3+\alpha T)T' r+\delta(T^4+\beta T^2)=0.
\label{eq:dif2}
\end{equation}
With $x=T^2$, choose explicit boundary data $x(r_0)=x_0=T_0^2$. The integrated equation can then be written in the dimensionless form
\begin{equation}
\left(\frac{x}{x_0}\right)^{\alpha/\beta}
\left(\frac{x+\beta}{x_0+\beta}\right)^{1-\alpha/\beta}
=\left(\frac r{r_0}\right)^{-2\delta/\gamma}.
\label{eq:sol2}
\end{equation}
For fixed positive boundary data, $T\to\infty$ as $r\to0^+$. In the high-temperature regime $T^2\gg\beta$,
\begin{equation}
T^2\propto r^{-2\delta/\gamma},
\qquad
\rho_{\rm cold\text{-}QGP}\propto r^{-4\delta/\gamma}.
\label{eq:cold-highT-scaling}
\end{equation}
In the formal low-temperature regime $T^2\ll\beta$, one obtains $T^2\propto r^{-8/3}$. The latter asymptote lies outside the intended inner high-density application and merely signals that the QGP branch must be matched to another outer phase.

\begin{figure}[tb]
\centering
\includegraphics[width=0.6\textwidth]{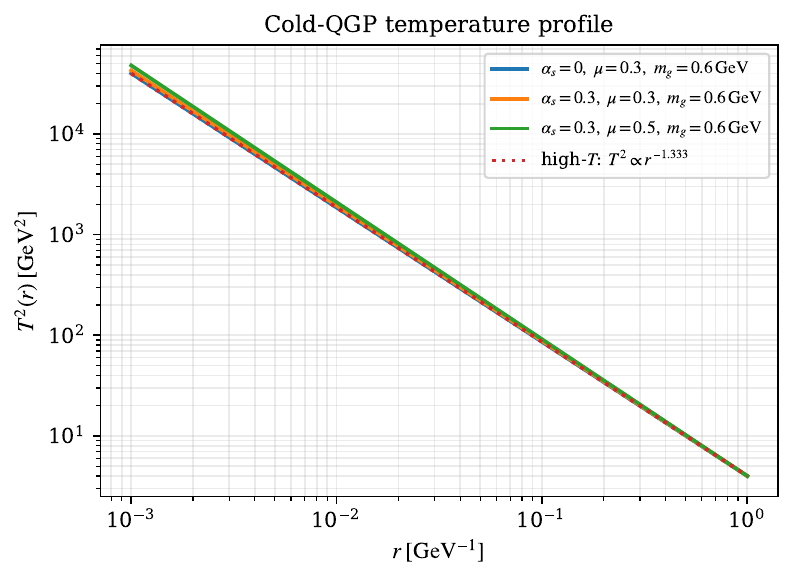}
\caption{Cold-QGP temperature profiles obtained by numerically inverting Eq.~\eqref{eq:sol2} with the common boundary condition $r_0=1~{\rm GeV}^{-1}$ and $T_0=2~{\rm GeV}$. The parameter triples $(\alpha_s,\mu,m_g)$ are $(0,0.30,0.60)$, $(0.3,0.30,0.60)$, and $(0.3,0.50,0.60)$, with energies in GeV. The dotted curve is the high-temperature asymptote for the first benchmark, $T^2\propto r^{-4/3}$. Because $T\geq2$~GeV over the plotted interval, the numerical curves remain within the intended high-temperature branch.}
\label{fig:coldQGP_T2}
\end{figure}

The mass equation gives
\begin{equation}
M'(r)=\frac12\rho_{\rm cold\text{-}QGP}r^2
\propto r^{2-4\delta/\gamma}.
\end{equation}
For $\delta/\gamma\neq3/4$,
\begin{equation}
M(v,r)\sim M_0(v)+D r^{3-4\delta/\gamma},
\qquad
D=\frac{C_\rho}{2(3-4\delta/\gamma)},
\label{eq:cold-mass-scaling}
\end{equation}
where $C_\rho>0$ is the leading density coefficient. At $\delta/\gamma=3/4$,
\begin{equation}
M-M_0\sim\frac{C_\rho}{2}\ln r.
\end{equation}
The definitions imply
\begin{equation}
\frac23\leq\frac{\delta}{\gamma}<1,
\qquad
-1<3-\frac{4\delta}{\gamma}\leq\frac13.
\end{equation}
When $\delta/\gamma<3/4$, $D>0$ and the correction vanishes as $r\to0$, but much more slowly than $r^3$. At $\delta/\gamma=3/4$ it is logarithmic, and for $\delta/\gamma>3/4$ the coefficient is negative and the formal function  diverges. In all cases the regular condition is violated.

\begin{figure}[tb]
\centering
\includegraphics[width=\textwidth]{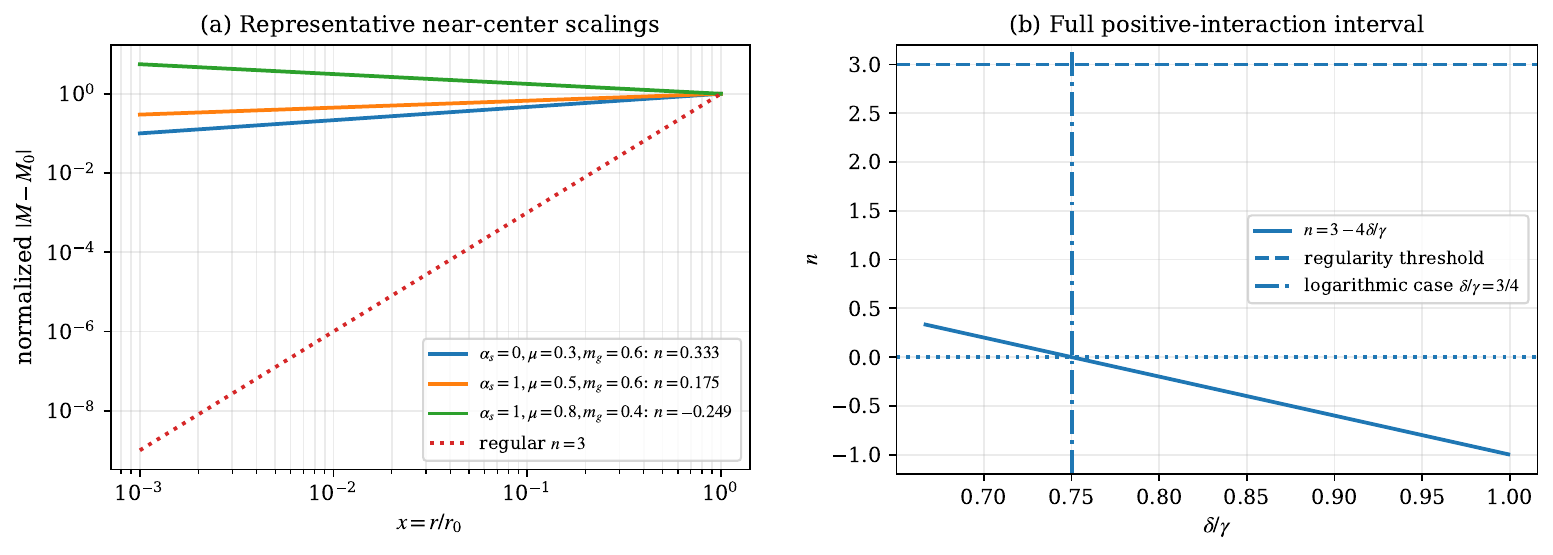}
\caption{Cold-QGP near-center mass scaling. Panel~(a): normalized $|M-M_0|\propto x^n$ for $(\alpha_s,\mu,m_g)=(0,0.30,0.60)$, $(1,0.50,0.60)$, and $(1,0.80,0.40)$, giving $n=0.333$, $0.175$, and $-0.249$, respectively; energies are in GeV. Panel~(b): $n=3-4\delta/\gamma$ over the complete positive-interaction interval. The vertical line $\delta/\gamma=3/4$ marks the logarithmic case, $n=0$, while the regularity threshold $n=3$ lies above the entire allowed interval.}
\label{fig:coldQGP_mass}
\end{figure}

\subsection{Curvature-based regularity criterion}

For a static metric function $F(r)=1-2M(r)/r$,
\begin{equation}
R=\frac{2(rM''+2M')}{r^2}.
\end{equation}
If $M\sim cr^n$, then
\begin{equation}
R\sim2cn(n+1)r^{n-3},
\end{equation}
and the exact leading Kretschmann coefficient is
\begin{equation}
K\sim4c^2\mathcal Q(n)r^{2n-6},
\qquad
\mathcal Q(n)=n^4-6n^3+17n^2-20n+12.
\label{eq:K-powerlaw-exact}
\end{equation}
Thus finite polynomial curvature invariants require $n\geq3$. The limiting case
\begin{equation}
M\sim\frac{\Lambda_{\rm eff}}6r^3
\end{equation}
gives $R\to4\Lambda_{\rm eff}$ and $K\to8\Lambda_{\rm eff}^2/3$.

\begin{figure}[tb]
\centering
\includegraphics[width=0.6\textwidth]{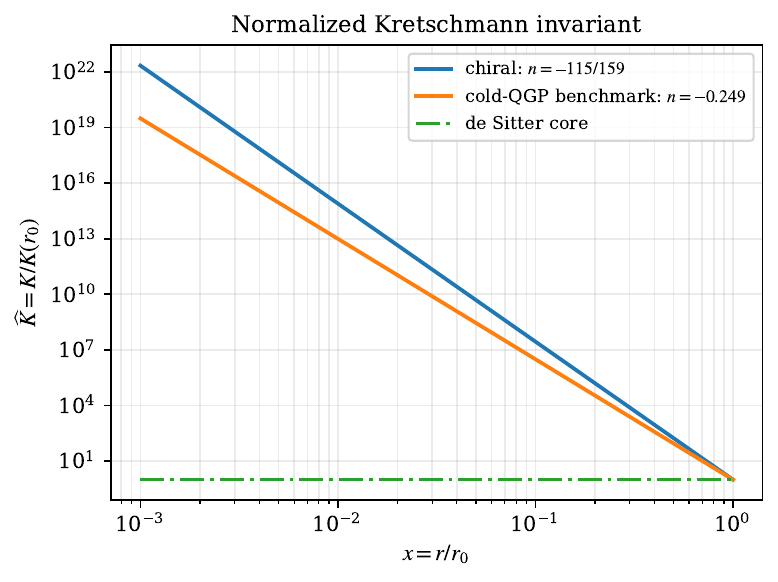}
\caption{Normalized Kretschmann invariant $\widehat K=K/K(r_0)$ calculated from the exact power-law coefficient in Eq.~\eqref{eq:K-powerlaw-exact}. The chiral curve uses $n=-115/159$; the cold-QGP curve uses the strong-interaction benchmark $(\alpha_s,\mu,m_g)=(1,0.80,0.40~{\rm GeV})$, for which $n=-0.249$. Both diverge as $r\to0$, whereas the de~Sitter curve is constant.}
\label{fig:Kretschmann}
\end{figure}

\subsection{General near-center no-go criterion}

The generalized Vaidya identity yields a model-independent criterion. Suppose that near the center
\begin{equation}
\frac{P_\perp}{\rho}\longrightarrow w_0,
\qquad \rho>0.
\end{equation}
Then
\begin{equation}
r\rho'+2(1+w_0)\rho\simeq0,
\end{equation}
so
\begin{equation}
\rho\propto r^{-2(1+w_0)},
\qquad
M\propto r^{1-2w_0},
\label{eq:general-scaling-theorem}
\end{equation}
except for the logarithmic case $w_0=1/2$. Curvature regularity requires $1-2w_0\geq3$, hence $w_0\leq-1$. The transverse NEC requires $\rho+P_\perp\geq0$, hence $w_0\geq-1$. Therefore, under the transverse NEC, the unique regular limiting behavior is
\begin{equation}
w_0=-1,
\qquad
P_\perp=-\rho,
\qquad
\rho\to\rho_0,
\qquad
M\propto r^3.
\label{eq:unique-regular-limit}
\end{equation}

The exact power-law exponents obtained for the two models depend on the adopted phenomenological pressure--density pairs. Nevertheless, the qualitative obstruction is more general. Any thermodynamically closed high-density equation of state whose effective transverse ratio approaches a finite value
\begin{equation}
w_0>-1
\end{equation}
gives
\begin{equation}
1-2w_0<3,
\end{equation}
and therefore fails the regular-center condition. Thus a change in the thermodynamic closure may alter the numerical exponent without removing the need for a vacuum-like limiting transverse stress.

For the chiral model, $w_0=137/159>0$. For the cold-QGP model,
\begin{equation}
w_0=\frac{37\pi^2/90+3\pi\alpha_s\mu^2/(4m_g^2)}{37\pi^2/30+3\pi\alpha_s\mu^2/(4m_g^2)}>0.
\end{equation}
Hence both equations of state fail this criterion independently of the detailed integration.

\subsection{Energy conditions and relation with the singularity theorems}

For the type-II source in Eq.~\eqref{eq:Tmunu-typeII}, and for the ingoing orientation in which $\sigma$ is the null-flux energy density, the standard pointwise conditions can be expressed as follows:
\begin{align}
\text{NEC:}\qquad&
\sigma\geq0,
\qquad
\rho+P_\perp\geq0,
\\
\text{WEC:}\qquad&
\sigma\geq0,
\qquad
\rho\geq0,
\qquad
\rho+P_\perp\geq0,
\\
\text{DEC:}\qquad&
\sigma\geq0,
\qquad
\rho\geq0,
\qquad
-\rho\leq P_\perp\leq\rho,
\\
\text{SEC:}\qquad&
\sigma\geq0,
\qquad
\rho+P_\perp\geq0,
\qquad
P_\perp\geq0.
\end{align}
The material radial principal combination is saturated identically,
\begin{equation}
\rho+P_\parallel=0,
\end{equation}
because
$P_\parallel=-\rho$.

For the chiral effective pair,
\begin{equation}
\rho_{\rm QGP}+P_\perp
=
\zeta T^4>0,
\end{equation}
while for the finite-density mean-field QGP pair,
\begin{equation}
\rho_{\rm cold\text{-}QGP}+P_\perp
=
\left(
\frac{74\pi^2}{45}
+
\frac{3\pi\alpha_s\mu^2}{2m_g^2}
\right)T^4
+
2\mu^2T^2
>0.
\end{equation}
In their leading high-temperature regimes, both models have
$\rho>0$
and
$0<P_\perp/\rho<1$.
Consequently, for a positive ingoing flux they satisfy the NEC, WEC, and SEC and asymptotically satisfy the DEC. Their singular behavior is therefore not produced by a violation of the null energy condition.

A de~Sitter core with
\begin{equation}
P_\parallel=P_\perp=-\rho
\end{equation}
saturates the principal null inequalities, satisfies the WEC and DEC, and violates the SEC. The classical singularity theorems also involve global and causal assumptions, including trapped surfaces and suitable causality or global-hyperbolicity hypotheses. An inner Cauchy horizon may evade some of these assumptions, but its presence is not by itself evidence of dynamical viability because such horizons can be vulnerable to blueshift and mass-inflation instabilities.

\subsection{Apparent horizons and possible naked singularities}

Marginally trapped spheres satisfy
\begin{equation}
F(v,r)=1-\frac{2M(v,r)}r=0.
\label{eq:horizon-condition}
\end{equation}
This condition locates apparent horizons but does not establish global censorship. Outgoing radial null geodesics satisfy
\begin{equation}
\frac{dr}{dv}=\frac12\left(1-\frac{2M(v,r)}r\right).
\label{eq:outgoing-null-geodesic}
\end{equation}
Local nakedness requires determining whether future-directed solutions of Eq.~\eqref{eq:outgoing-null-geodesic} can emerge from the singular boundary. That question depends on the full $v$ dependence, integration constants, initial data, and exterior matching, none of which are fixed by the quasi-local radial reconstruction alone. We therefore do not automatically call the singular QCD branches black holes. They are interpreted as black-hole interiors only when a trapped region exists and the global completion hides the singular boundary; otherwise they remain singular collapse solutions that may be locally naked. The regular matched examples below are restricted to parameter sets with at least one positive apparent-horizon root, without claiming that this alone proves global cosmic censorship.

\section{Transition of baryonic matter into quark matter}
\label{sec:transition}

The QCD branches studied above describe local high-density phases but do not constitute a dynamical simulation of conversion from baryonic matter. We therefore formulate the transition as a finite-time phenomenological source model and state explicitly what it can and cannot accomplish.

The conversion begins at spacetime points where the local state crosses the adopted phase boundary,
\begin{equation}
T(v,r)=T_c[\mu(v,r)].
\label{eq:transition-threshold}
\end{equation}
After a definite and monotonic equation of state has been specified, the same phase boundary may alternatively be expressed as a critical-density condition,
\begin{equation}
\rho_b(v,r)=\rho_c[\mu(v,r)].
\end{equation}
These two expressions are not automatically equivalent without an explicit thermodynamic map between
$T$, $\mu$, and $\rho_b$.

A microscopic conversion rate
$\Gamma_{\rm tr}$
is defined per unit comoving proper time. The characteristic proper conversion interval is therefore
\begin{equation}
\Delta\tau_{\rm tr}
\sim
\Gamma_{\rm tr}^{-1}.
\end{equation}
Along a fluid worldline, the corresponding coordinate-time and coordinate-radial intervals are
\begin{equation}
\Delta v_{\rm tr}
\sim
u^v\Delta\tau_{\rm tr},
\qquad
\Delta r_{\rm tr}
\sim
|u^r|\Delta\tau_{\rm tr}.
\label{eq:transition-times}
\end{equation}
These coordinate intervals are gauge dependent; the invariant microscopic quantity is
$\Delta\tau_{\rm tr}$.

The conversion time should be compared with the local dynamical scale
\begin{equation}
t_{\rm dyn}\sim\frac{1}{\sqrt{G\rho}}.
\end{equation}
For reference, a homogeneous Newtonian sphere has the free-fall estimate
\begin{equation}
t_{\rm ff}
=
\sqrt{\frac{3\pi}{32G\rho}}.
\end{equation}
The present radial equations are quasi-local snapshots of a finite transition layer. They cannot predict a numerical onset time without a dynamical solution for
$T(v,r)$,
$\mu(v,r)$,
$u^\mu$,
and the microscopic rate
$\Gamma_{\rm tr}(T,\mu)$.

In order to estimate the time of conversion, one needs to diagonalize the line element. In its current form, due to off-diagonal term, the Eddington time $v$ is not measurable. However, the method of diagonalization of this line element is possible analyticaly for very narrow classes of mass functions. For considered models it is possible to do only numerically.

A fully covariant interaction between the two components would be described by a source four-vector $Q^\nu$,
\begin{equation}
\nabla_\mu T_b^{\mu\nu}=-Q^\nu,
\qquad
\nabla_\mu T_{\rm rad}^{\mu\nu}=Q^\nu,
\label{eq:covariant-conversion-source}
\end{equation}
so that the total stress tensor is conserved. The following radial equations should be understood as a quasi-local phenomenological projection of such a covariant system:
\begin{align}
r\rho_{\rm rad}'
+
2(\rho_{\rm rad}+P_{\rm rad})
&=
\mathcal Q(v,r),
\\
r\rho_b'
+
2(\rho_b+P_b)
&=
-\mathcal Q(v,r).
\label{eq:conversion-system}
\end{align}
A seed-independent radial closure is
\begin{equation}
\mathcal Q(v,r)
=
\beta(v,r)\rho_b(v,r),
\label{eq:conversion-source}
\end{equation}
where $\beta(v,r)$ is a dimensionless phenomenological transfer function that vanishes before the transition threshold is crossed. Unlike a source proportional to
$\rho_{\rm rad}$,
this closure can initiate conversion from a state with no pre-existing radiation component.

No universal relation between
$\beta$
and the proper-time rate
$\Gamma_{\rm tr}$
can be inferred from the radial equations alone. Such a relation depends on the covariant source vector
$Q^\mu$,
the fluid four-velocity, the direction of the flow, and the chosen radial coordinate.

For a prescribed transfer function $\beta(v,r)$, the baryonic density is
\begin{equation}
\rho_b(v,r)
=
C_b(v)r^{-2(1+\alpha)}
\exp\!\left[
-\int^r
\frac{\beta(v,\bar r)}{\bar r}
\,d\bar r
\right],
\label{eq:rho-b-solution}
\end{equation}
and the radiation density is
\begin{equation}
\rho_{\rm rad}(v,r)
=
r^{-8/3}
\left[
C_{\rm rad}(v)
+
\int^r
\beta(v,\bar r)
\rho_b(v,\bar r)
\bar r^{5/3}
\,d\bar r
\right].
\label{eq:rho-rad-solution}
\end{equation}
These expressions describe a phenomenological radial redistribution between two positive-pressure components. They neither determine the microscopic QCD conversion rate nor replace a covariant Einstein--hydrodynamic evolution.

There is a decisive limitation. For $\rho_b\geq0$, $\rho_{\rm rad}\geq0$, and $\alpha\geq0$,
\begin{equation}
P_{\rm mix}=\alpha\rho_b+\frac13\rho_{\rm rad}\geq0,
\qquad
\rho_{\rm mix}=\rho_b+\rho_{\rm rad}>0,
\end{equation}
and therefore
\begin{equation}
\min\!\left(\alpha,\frac13\right)
\leq\frac{P_{\rm mix}}{\rho_{\rm mix}}
\leq\max\!\left(\alpha,\frac13\right).
\label{eq:mixture-bound}
\end{equation}
A positive-energy baryon--radiation mixture can never approach $P=-\rho$. The conversion process can change the QCD composition and redistribute internal energy, but it cannot by itself generate a de~Sitter core. Such a core requires a separate vacuum component, condensate, nonperturbative phase, or other sector with $P_{\rm vac}\simeq-\rho_{\rm vac}$.

Radiation produced after an event horizon has formed cannot escape to future null infinity. It remains part of the interior stress tensor and its backreaction must be included. Consequently, radiation emission cannot be invoked as a mechanism that simply carries away the energy responsible for the central singularity. In the following section we adopt the conservative interpretation: the QCD conversion determines an outer high-density layer, while an additional explicitly vacuum-like component supplies the regular inner completion.

\section{Explicit matching to an inner vacuum-like core}
\label{sec:corematching}

We now construct an effective completion in which the singular QCD branch is replaced below a transition radius by a vacuum-like core. This construction is not derived from the QCD equations of state; it displays the minimal additional structure required by the no-go result.

Regularity requires
\begin{equation}
M(v,r)=\frac{\Lambda_{\rm eff}(v)}6r^3+\mathcal O(r^4),
\qquad r\to0,
\label{eq:regularityM}
\end{equation}
so
\begin{equation}
M(v,0)=M'(v,0)=M''(v,0)=0.
\label{eq:centerconditions}
\end{equation}
Using Eq.~\eqref{eq:stress-identification},
\begin{equation}
\rho(0)=\Lambda_{\rm eff},
\qquad
P_\parallel(0)=P_\perp(0)=-\Lambda_{\rm eff}.
\end{equation}
A switching function that is merely exponentially small is insufficient to eliminate an exact power-law singularity. We therefore use a compact-support $C^\infty$ function.

Define
\begin{equation}
\chi(x)=\begin{cases}e^{-1/x},&x>0,\\0,&x\leq0,\end{cases}
\qquad
s(r)=\frac{r-(r_c-\Delta)}{2\Delta},
\end{equation}
and
\begin{equation}
S(r)=\frac{\chi[s(r)]}{\chi[s(r)]+\chi[1-s(r)]}.
\label{eq:finterp}
\end{equation}
Then
\begin{equation}
S(r)=0\quad(r\leq r_c-\Delta),
\qquad
S(r)=1\quad(r\geq r_c+\Delta).
\end{equation}
The matched mass function is
\begin{equation}
M_{\rm eff}(v,r)=S(r)M_{\rm QCD}(v,r)+[1-S(r)]\frac{\Lambda_{\rm eff}(v)}6r^3.
\label{eq:Meff}
\end{equation}
The singular QCD branch is therefore exactly absent from the inner core and exactly recovered outside the transition layer.

In the explicit construction below, the matching radius $r_c$ and thickness $\Delta$ are held fixed, so that
\begin{equation}
S=S(r).
\end{equation}
The construction should therefore be interpreted as static or quasi-static. If the transition layer evolves, one must instead use
$S=S(v,r)$.
In that case,
$\partial_v S$
produces additional contributions to
\begin{equation}
\sigma=\frac{2\dot M_{\rm eff}}{r^2},
\end{equation}
and these terms must be included in the flux and energy-condition analysis.

The effective density is
\begin{align}
\rho_{\rm eff}
&=\frac{2}{r^2}\partial_rM_{\rm eff}\\
&=S\rho_{\rm QCD}+(1-S)\Lambda_{\rm eff}
+\frac{2S'}{r^2}\left(M_{\rm QCD}-\frac{\Lambda_{\rm eff}}6r^3\right).
\label{eq:rhoeff2}
\end{align}
The radial principal pressure remains fixed by the geometry,
\begin{equation}
P_{\parallel,{\rm eff}}=-\rho_{\rm eff}.
\label{eq:Prad-eff}
\end{equation}
The transverse pressure is
\begin{align}
P_{\perp,{\rm eff}}
&=-\frac1r\partial_r^2M_{\rm eff}\\
&=S P_{\perp,{\rm QCD}}-(1-S)\Lambda_{\rm eff}
-\frac{2S'}r\partial_rM_{\rm QCD}
-\frac{S''}r\left(M_{\rm QCD}-\frac{\Lambda_{\rm eff}}6r^3\right)
+\Lambda_{\rm eff}rS'.
\label{eq:Peff2}
\end{align}
The $S'$ and $S''$ terms are localized stresses of the matching layer. The radial NEC is saturated identically, whereas the transverse NEC is controlled by $\rho_{\rm eff}+P_{\perp,{\rm eff}}$ and may be violated locally in the layer. The core itself saturates both principal null conditions and violates the strong energy condition.

\begin{figure}[tb]
\centering
\includegraphics[width=\textwidth]{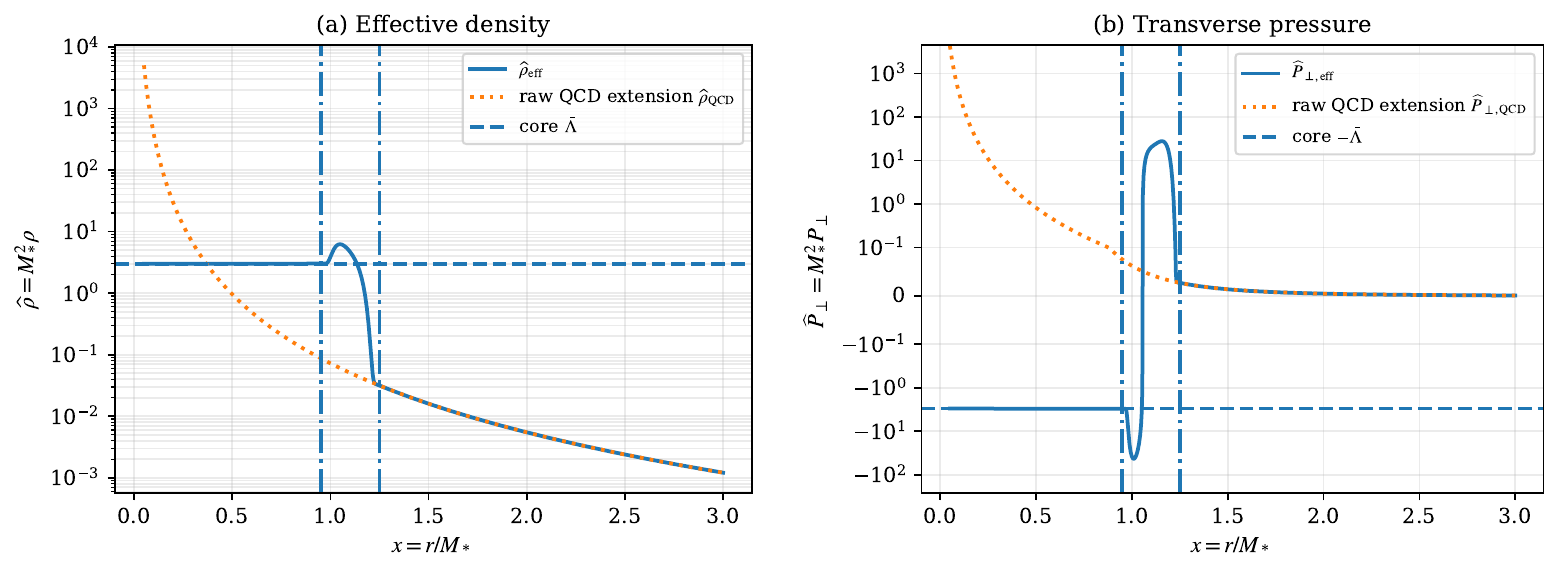}
\caption{Dimensionless effective density and transverse pressure for the benchmark in Eqs.~\eqref{eq:benchmark-outer-mass}--\eqref{eq:benchmark-matching-parameters}. Panel~(a): $\widehat\rho=M_*^2\rho$ remains nonnegative and approaches $\overline\Lambda=3$ in the core. Panel~(b): $\widehat P_\perp=M_*^2P_\perp$ approaches $-3$ in the core and develops localized positive and negative stresses in the transition layer $0.95<x<1.25$. The dotted curves show the unmodified singular QCD extension. The radial pressure is not shown because $P_\parallel=-\rho$ identically.}
\label{fig:effective_fluid}
\end{figure}

\begin{figure}[tb]
\centering
\includegraphics[width=\textwidth]{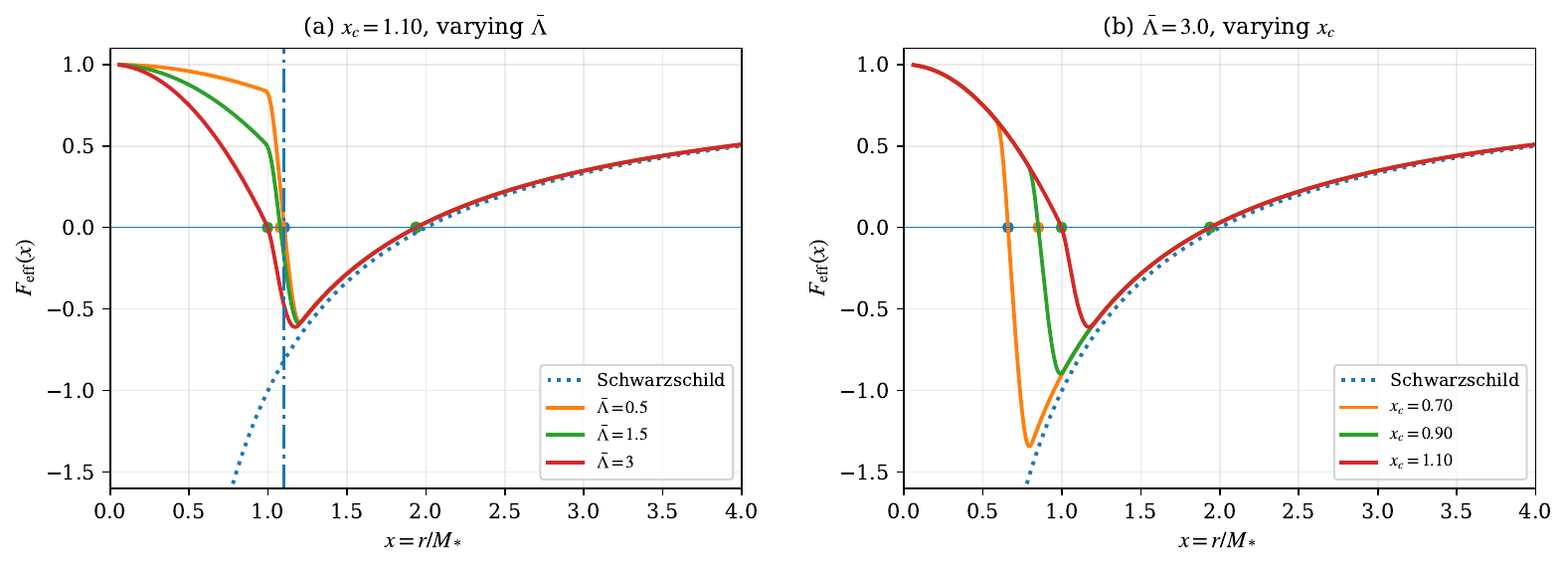}
\caption{Effective lapse for the explicit outer profile in Eq.~\eqref{eq:benchmark-outer-mass}, with $\Delta_x=0.15$. Filled circles mark apparent-horizon roots. Panel~(a): fixed $x_c=1.10$ and $\overline\Lambda=0.5,1.5,3.0$. Panel~(b): fixed $\overline\Lambda=3.0$ and $x_c=0.70,0.90,1.10$. The common outer root is $x_+\simeq1.938$, rather than exactly $2$, because the QCD correction has not vanished completely there. Every matched profile satisfies $F_{\rm eff}(0)=1$.}
\label{fig:lapse}
\end{figure}

\begin{figure}[tb]
\centering
\includegraphics[width=0.6\textwidth]{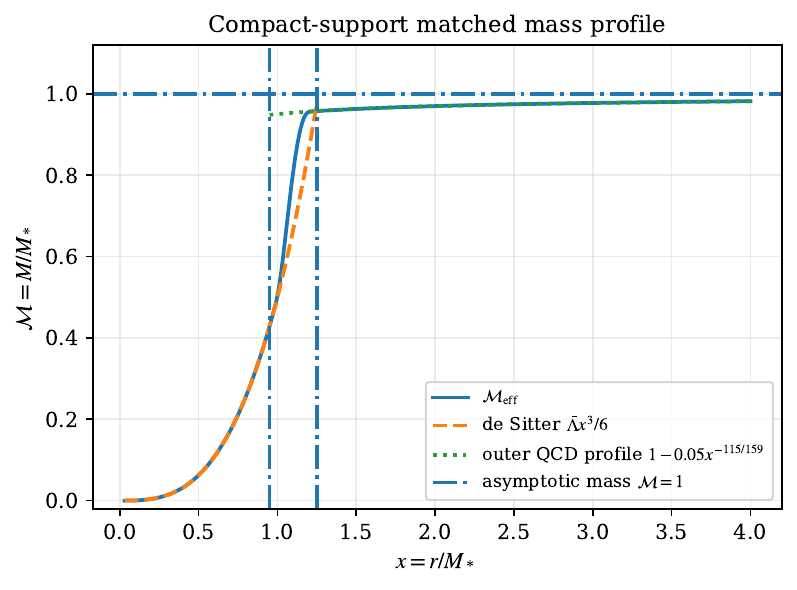}
\caption{Matched dimensionless mass $\mathcal M_{\rm eff}$ in Eqs.~\eqref{eq:benchmark-outer-mass}--\eqref{eq:benchmark-matching-parameters}. The profile equals $\overline\Lambda x^3/6$ for $x\leq0.95$, equals $1-0.05x^{-115/159}$ for $x\geq1.25$, and tends to the asymptotic mass $\mathcal M=1$. The vertical lines delimit the compact-support transition layer.}
\label{fig:matched_mass}
\end{figure}

For a fully reproducible numerical illustration, Figs.~\ref{fig:effective_fluid}--\ref{fig:matched_mass} use one common dimensionless benchmark. We define
\begin{equation}
x\equiv\frac r{M_*},
\qquad
\mathcal M\equiv\frac M{M_*},
\qquad
\overline\Lambda\equiv\Lambda_{\rm eff}M_*^2,
\end{equation}
and choose the outer chiral scaling
\begin{equation}
\mathcal M_{\rm QCD}(x)=1-0.05\,x^{-115/159}.
\label{eq:benchmark-outer-mass}
\end{equation}
This profile has positive density, tends to the asymptotic mass $M_*$ as $x\to\infty$, and is finite throughout the matching layer, although it is singular if extrapolated to $x=0$. The reference matched configuration uses
\begin{equation}
x_c=1.10,
\qquad
\Delta_x=0.15,
\qquad
\overline\Lambda=3.0.
\label{eq:benchmark-matching-parameters}
\end{equation}

Because $S$ is $C^\infty$ and constant outside the transition layer, $M_{\rm eff}$ and all of its derivatives match smoothly provided $M_{\rm QCD}$ is finite on $[r_c-\Delta,r_c+\Delta]$. No evaluation of the singular QCD branch at $r=0$ is required. For the benchmark of Fig.~\ref{fig:effective_fluid}, the density remains positive throughout the plotted domain, while $\rho_{\rm eff}+P_{\perp,{\rm eff}}$ becomes negative in part of the transition layer. Thus the explicit regular completion preserves positive energy density but locally violates the transverse NEC, as expected for a nontrivial interpolation between the QCD branch and the de~Sitter core.

The effective lapse is denoted by
\begin{equation}
F_{\rm eff}(v,r)=1-\frac{2M_{\rm eff}(v,r)}r,
\label{eq:Feff}
\end{equation}
to avoid confusing it with the switching function $S(r)$. Apparent horizons are positive roots of $F_{\rm eff}=0$. Depending on $(r_c,\Delta,\Lambda_{\rm eff})$ and the outer mass profile, the composite geometry can possess two, one, or no apparent horizons. A double root is the extremal boundary between a regular black hole and a horizonless compact object.

For the reference configuration $(x_c,\Delta_x,\overline\Lambda)=(1.10,0.15,3.0)$, the numerical roots are
\begin{equation}
x_-\simeq0.9974,
\qquad
x_+\simeq1.9380.
\end{equation}
At fixed $x_c=1.10$, increasing $\overline\Lambda$ from $0.5$ to $3.0$ moves the inner root from approximately $1.1034$ to $0.9974$, while the outer root remains fixed by the common QCD-supported exterior. At fixed $\overline\Lambda=3.0$, changing $x_c$ from $0.70$ to $1.10$ moves the inner root from approximately $0.6592$ to $0.9974$. These values are illustrative and are not predictions of the QCD equations of state; they quantify the chosen effective completion.

A viable parameter set must satisfy: (i) regularity at the center; (ii) nonnegative density in the matching layer; (iii) any desired energy-condition restrictions; and (iv) the required apparent-horizon structure. The curvature invariants approach
\begin{equation}
R\to4\Lambda_{\rm eff},
\qquad
K\to\frac83\Lambda_{\rm eff}^2
\end{equation}
as $r\to0$.

The matching construction should therefore be interpreted as a controlled effective completion, not as evidence that QCD conversion dynamically produces vacuum energy. Its phenomenological parameters can be constrained only after a microscopic inner phase is specified. Once a consistent profile is available, its horizons, photon spheres, linear perturbations, and observational signatures can be studied without extrapolating the singular QCD branch into the core.

\section{Conclusion}
\label{sec:conclusion}

We have investigated whether two finite-density QCD-inspired effective equations of state can generate a regular center within a generalized Vaidya collapse geometry. The scope of the result is important. The generalized Vaidya Einstein tensor fixes the radial principal pressure to
$P_\parallel=-\rho$.
We therefore adopted the phenomenological anisotropic closure
$P_\perp=p_{\rm QCD}$,
with
$P_\perp=-M''/r$.
The conclusions apply to this geometrical reconstruction and should not be interpreted as a general no-go theorem for all dynamical configurations of QCD matter.

For the chiral effective pair, the physical principal Lambert-$W$ branch gives
\begin{equation}
T^2\propto r^{-296/159},
\end{equation}
leading to a non-integrable central density and a singular mass contribution
$M-M_0\propto r^{-115/159}$.
For the finite-density mean-field QGP pair, the implicit solution gives
\begin{equation}
T^2\propto r^{-2\delta/\gamma},
\qquad
M-M_0\propto r^{3-4\delta/\gamma},
\end{equation}
with a mass exponent strictly smaller than three throughout the physical parameter interval of the adopted approximation. The precise numerical exponents are properties of the phenomenological pressure--density pairs and may change in a fully thermodynamically closed treatment.

The qualitative result is more general. If the effective transverse equation of state approaches
\begin{equation}
\frac{P_\perp}{\rho}\longrightarrow w_0,
\end{equation}
the generalized Vaidya identity gives
\begin{equation}
\rho\propto r^{-2(1+w_0)},
\qquad
M\propto r^{1-2w_0}.
\end{equation}
Regularity requires
$w_0\leq-1$,
whereas the transverse null energy condition requires
$w_0\geq-1$.
Under the NEC, the unique regular limiting behavior is therefore
\begin{equation}
w_0=-1,
\qquad
P_\perp=-\rho,
\qquad
M\propto r^3.
\end{equation}
Neither effective QCD closure studied here approaches this vacuum-like transverse limit.

We have also clarified the limitations of the phase-transition model. A microscopic conversion time must be defined in comoving proper time. A density threshold is equivalent to a temperature threshold only after an explicit thermodynamic map has been specified. Moreover, the radial exchange equations are phenomenological projections of a covariant source system and do not determine a microscopic conversion rate. A positive-energy mixture of baryonic matter and radiation cannot generate
$P=-\rho$,
and radiation produced after horizon formation cannot simply remove the energy responsible for the central singularity.

Finally, we constructed a compact-support $C^\infty$ interpolation to an inner de~Sitter-like component. This construction removes the singular QCD extension exactly from the core and illustrates the additional vacuum-like physics required by the conditional no-go result. It is an effective static or quasi-static completion rather than a first-principles consequence of QCD conversion. A complete treatment will require a thermodynamically consistent finite-density equation of state, a covariant microscopic conversion source, dynamical stress anisotropy, and a coupled Einstein--hydrodynamic evolution. The stability of any inner Cauchy horizon and the possible effects of mass inflation must also be investigated before such a regular completion can be regarded as dynamically viable.

\acknowledgments

A.{\"O}. and G.L.\ would like to acknowledge the contribution of the COST Action CA21106 -- COSMIC WISPers in the Dark Universe: Theory, astrophysics and experiments (CosmicWISPers), the COST Action CA21136 -- Addressing observational tensions in cosmology with systematics and fundamental physics (CosmoVerse), the COST Action CA22113 -- Fundamental challenges in theoretical physics (THEORY-CHALLENGES), the COST Action CA23130 -- Bridging high and low energies in search of quantum gravity (BridgeQG) and the COST Action CA23115 -- Relativistic Quantum Information (RQI) funded by COST (European Cooperation in Science and Technology). We also thank EMU, TUBITAK, ULAKBIM (T\"urkiye) and SCOAP3 (Switzerland) for their support.

\bibliographystyle{apsrev4-1}
\bibliography{ref}

\end{document}